# DEVELOPMENT OF MECHANISM FOR ENHANCING DATA SECURITY IN QUANTUM CRYPTOGRAPHY.

Ajit Singh[1] and Nidhi Sharma[2]

Department of Computer Science & Engineering and Information Technology
BPS Mahila Vishwavidyalaya, Khanpur Kalan, Sonepat-131305 Haryana (India).
[1]ghanghas_ajit@rediffmail.com
[2]nidhi.sharma.1012@gmail.com

## ABSTRACT

*Nowadays security in communication is increasingly important to the network communication because many categories of data are required restriction on authorization of access, modify, delete and insert. Quantum cryptography is one of the solutions that use property of polarization to ensure that transmitted data is not tampered. The research paper provides the mechanism that enhances the data security in quantum cryptography during exchange of information. In first phase detailed explanation of Quantum key distribution's BB84 protocol is given. BB84 protocol is used as the basis for the mechanism. In next phase the proposed mechanism is explained. The proposed mechanism combines BB84 protocol at two levels, from sender to receiver and then from receiver to sender. Moreover, a logic circuit is used to combine the bits hence to reduce the probability of eavesdropping. The key obtained can be used to exchange the information securely further it can help in encryption and decryption of crucial data. Double level BB84 mechanism will help in information reconciliation as well as privacy amplification. In future the proposed mechanism will be very beneficial where unconditional security is required during key and other secret information exchange*

## 1. INTRODUCTION

Quantum Cryptography is composed of two words: Quantum and Cryptography. Quantum is the smallest discrete quantity of some physical property that a system can possess and Cryptography enables to store sensitive information or transmit it across insecure networks so that it cannot be read by anyone except the intended recipient. So, Quantum Cryptography is using the quantum for doing cryptographic tasks. Quantum Cryptography is based upon conventional cryptographic methods and extends these through the use of quantum effects. [16] Quantum Key Distribution (QKD) is used in quantum cryptography for generating a secret key shared between two parties using a quantum channel and an authenticated classical channel. The private key obtained then used to encrypt messages that are sent over an insecure classical channel (such as a conventional internet connection).

In traditional Cryptography the security depends upon how complex a mathematical problem is to solve. In today's scenario with the advent of technologies these complex mathematical problems are no longer hard to solve. Hence security level reduces. Modern cryptosystem uses Quantum Cryptography that makes the key unconditionally secure with quantum mechanics. For example: Heisenberg's Uncertainty Principle, Wave/Particle duality, Qubits and No cloning Theorem. Heisenberg's Uncertainty principle states that the more precisely one property is measured, the less precisely the other can be measured. [1,17] Using this principle Quantum Cryptography successfully provides unconditional security. [2] The concept of Wave/Particle

Duality is being used in photon polarization. A qubit or quantum bit is a unit of quantum information. Like a bit a qubit can have values 0 or 1, a qubit can retain superposition state of these two bits. The no cloning theorem implies that a possible eavesdropper can not intercept, measure and reemit a photon without introducing a significant and detectable error in the reemitted signal. Thus, it is possible to build a system that allows two parties, the sender and the receiver, commonly called "Alice" and "Bob", to exchange information and detect where the communication channel has been tempered with. [3] The key obtained using quantum cryptography can then be used with any chosen encryption algorithm to encrypt (and decrypt) a message, which can be transmitted over a standard communication channel. Once the secret key using Quantum Cryptography is established, it can be used together with classical cryptographic techniques such as the one-time-pad to allow the parties to communicate meaningful information in absolute secrecy. [4]

## 2. QUANTUM CRYPTOGRAPHY

In QKD, two parties, Alice and Bob, obtain some quantum states and measure them. A QKD system consists of a quantum channel and a classical channel. The quantum channel is only used to transmit Qubits (single photons) and must consist of a transparent optical path. The classical channel can be a conventional IP channel. The Key generation in QKD is done by communicating through quantum channels [9].They communicate through classical channel to determine which of their measurement results could lead to secret key bits. QKD [5, 6, 7] systems continually and randomly generate new private keys that both parties share automatically. A compromised key in a QKD system can only decrypt a small amount of encoded information because the private key may be changed every second or even continuously. To build up a secret key from a stream of single photons, each photon is encoded with a bit value of 0 or 1, typically by a photon in some superposition state, such as polarization. These photons are emitted by a conventional laser as pulses of light so dim that most pulses do not emit a photon. [8] This way of communication has the ability to create true random and secret key, which can then be used as seeds to conventional cryptographic methods for the generation of suitable keys. The exchange of information through Quantum channel can be shown with the help of diagram [10].

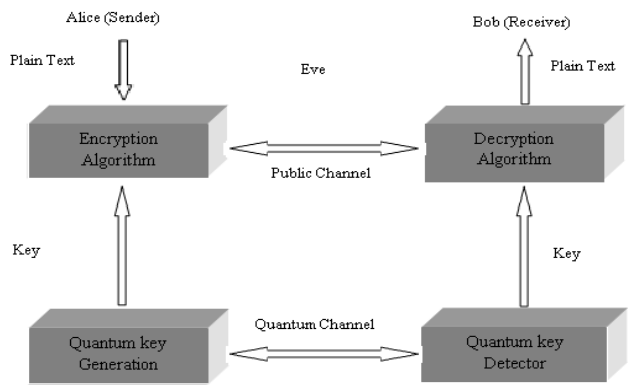

Figure 1. Quantum Key Distribution

### 2.1 BB84 Protocol

BB84 is the first quantum cryptographic protocol that was introduced in 1984 by Charles H. Bennet of IBM New York and Gilles Brassard of the University of Montreal. In opposition to public key systems this protocol is based upon the generation of random secret (private) encryption and decryption keys. It allows two users to establish an identical and purely random sequence of bits while allowing revealing of any eavesdropping. BB84 uses two pairs of states, with each pair conjugate to the other pair, and the two states within a pair orthogonal to each other. Pairs of orthogonal states are referred to as a basis. The usual polarization state pairs used are either the rectilinear basis of vertical (0°) and horizontal (90°), the diagonal basis of 45° and 135° or the circular basis of left- and right-handedness.

| Basis | 0 | 1 |
|---|---|---|
| + | → | ↑ |
| X | ↙ | ↗ |

Figure 2. Polarization States

In the lab experiment [11], the BB84 protocol encodes single photon polarizations using two bases of the same 2–dimensional Hilbert space. Only requirement on the involved quantum states is actually that they belong to mutually non-orthogonal bases of their Hilbert space. If a measurement on a system is performed in a basis different from the one the system is prepared in, its outcome is completely random and the system looses all the memory of its previous state. On the other hand, measurements performed in the basis identical to the basis of preparation of states will produce deterministic results. The protocol relies on Heisenberg's uncertainty principle, which forbids the measurement of more than one polarization component of one photon. BB84 is explained in the Figure 3. [14]

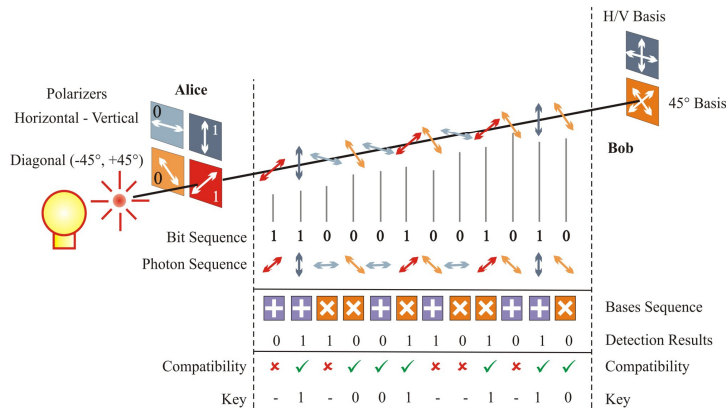

Figure3. The Quantum Key Distribution System using BB84 Protocol

To exchange a secret key in the BB84 protocol [12], Alice and Bob must do as follow:

- Alice creates a binary random number and sends it to Bob using randomly the two different bases + (rectilinear) and X (diagonal). Therefore, Alice transmits photons randomly in the four polarization states:

$$|\rightarrow\rangle, |\uparrow\rangle, |\nearrow\rangle, \text{ and } |\searrow\rangle.$$

- Bob simultaneously measures the polarization of the incoming photons using randomly the two different bases. He does not know which of his measurements are deterministic, i.e. measured in the same basis as the one used by Alice.

- Later, Alice and Bob communicate to each other the list of the bases they used. This communication carries no information about the value of the measurement, but allows Alice and Bob to know which values were measured by Bob correctly.

- Bob and Alice keep only those bits that were measured deterministically and will disregard those sent and measured in different bases. Statistically, their bases coincide in 50% of all cases, and Bob's measurements agree with Alice's bits perfectly.

- Together, they can reconstitute the random bit string created previously by Alice.

## 3. QUANTUM KEY DISTRIBUTION USING BB84 AND B92 PROTOCOL

Ching-Nung Yang and Chen-Chin Kuo, in their research paper [15] combined the BB84 protocol and B92 protocols and B92 and B92 protocols twice for improving the efficiency and performance. A brief description of their research work is given as follows:

In their research paper they introduced two new enhanced protocols:

- FEQKD i.e. First Enhanced Quantum Key Distribution protocol in which one four state protocol BB84 protocol and the other two state protocol B92 protocol is combined (BB84 + B92).

- SEQKD i.e. Second Enhanced Quantum Key Distribution protocol in which both two state protocols i.e. B92 is combined with B92 protocol during transmission from Alice to Bob and then from Bob to Alice.

They calculated the idealized maximum efficiency 42.9% and the complexity order 2.86 for FEQKD. It has better efficiency and a little complexity than B92 protocol, but when compared with BB84 protocol it has simpler complexity and a little less efficiency. For SEQKD protocol they used B92 protocol and were successful in enhancing the efficiency for B92 protocol by adding extra steps. For FEQKD and SEQKD protocols they use the information when Bob chooses the wrong detector's basis; however the information is discarded in original BB84 protocol.

We have forwarded this concept by combining BB84 protocol twice along with the logic circuit to enhance the security level.

## 4. PROPOSED TECHNIQUE

In proposed technique we use BB84 protocol as the base. The photons are sent in combination of two bits. A logic circuit is applied which is given below in figure 3 In logical circuit If the both bits are in the state either 1,or 0 then the output remains identical otherwise swapping is done between two bits. The truth table and the realization of the logical circuit are given in table 1 and Figure 4. The bits are sent through this logic circuit and then passed through the polarizer.

| Input | | Output | |
|---|---|---|---|
| A | B | A | B |
| 0 | 0 | 0 | 0 |
| 0 | 1 | 1 | 0 |
| 1 | 0 | 0 | 1 |
| 1 | 1 | 1 | 1 |

Table 1. Truth Table

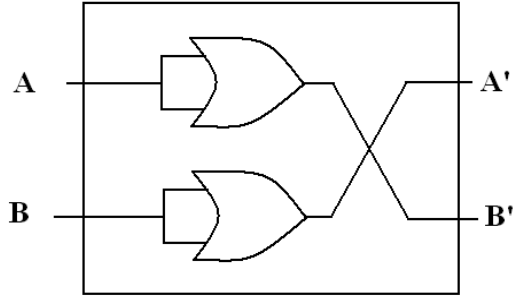

Figure 3. Realization of logic circuit

### 4.1 Steps in Proposed Technique

In 1st stage the transmission is done from Alice to Bob.

- Alice creates a binary sequence that is to be sent to Bob.
- In 2nd step the bits are passed through the defined logic circuit in. The information about the logic circuit is already exchanged between Alice and Bob.
- The output of logic circuit is sent to Bob using randomly the two different bases + (rectilinear) and X (diagonal)

$|\uparrow\rangle, |\nearrow\rangle$ both represent 1

$|\rightarrow\rangle, |\nwarrow\rangle$ both represents 0

Therefore, Alice transmits photons randomly in the four polarization states

$|\uparrow\rangle, |\rightarrow\rangle, |\nwarrow\rangle$ and $|\nearrow\rangle$

- Bob randomly chooses the basis and the polarization is done.
- Later, Alice and Bob communicate to each other the list of the bases they used. This communication carries no information about the value of the measurement, but allows Alice and Bob to know which values was measured by Bob correctly. On the deterministic bits obtained at Bob end the logic circuit is applied.
- Bob and Alice keep only those bits that were measured deterministically and will disregard those sent and measured in different bases.

Here 1st stage of BB84 is completed.

Now the **2nd stage** starts. In which transmission is done from Bob to Alice. Basically in 2nd stage BB84 protocol is applied for Information reconciliation. In 2nd stage I have tried to increase this number so that accuracy relative to the original bit sequence decided by Alice can be achieved in a better way.

- Bob decides the Source Basis only for those bits that did not match and applies the polarization.
- Alice puts her predefined basis for measurement and applies the polarization.
- Then Alice does measurement of bit combination that is common.

These combinations of bits are further added to the 1st stage shared keys. The final output bits are the set of completely secure key. Out of these bits any subset can be taken as final secret quantum key.

The flow chart regarding this technique is given as below in figure 4.

## 4.2 Implementation

The demonstration of the given technique with a sample example and steps that take place in the transmission from Alice to Bob and Bob to Alice is explained in this section.

### 4.2.1 First Stage from Alice to Bob

1. Alice decided the binary sequence of 16 bits 1100011010010011 for secret key.
2. In 2nd step the bits are passed through the defined logic circuit in figure 3. After applying the logic circuit the output is: 1100100101100011

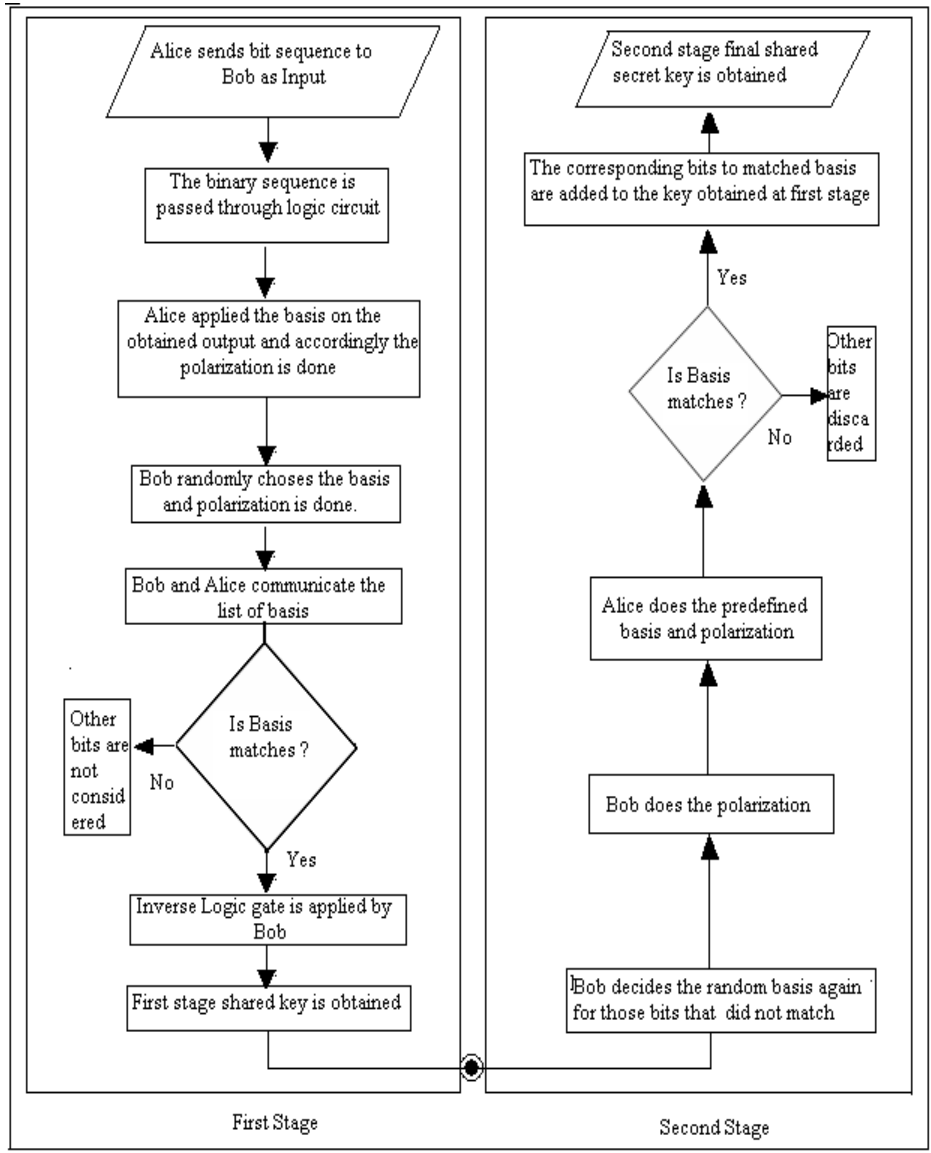

Figure 4. Flow Chart of First and Second Stage Transmission

3. The output of logic circuit is sent to Bob using randomly the two different bases + (rectilinear) and X (diagonal)
   Therefore, Alice transmits photons randomly in the four polarization states. The Basis decided for the given output are X++X++XXXX+++XX+. Accordingly the polarization is done for this basis at sender side.
4. Bob randomly chooses the basis and the polarization is done. The random basis chosen by Bob are: +++XX+XXXXX++X++. On this random chosen basis the polarization is performed with the help of polarizer.
5. Later, Alice and Bob communicate to each other the list of the bases they used. This communication carries no information about the value of the measurement, but allows Alice and Bob to know which values were measured by Bob correctly. On the deterministic bits obtained at Bob end the logic circuit is applied. Also, Bob apply the inverse logic gate.
6. Bob and Alice keep only those bits that were measured deterministically and will disregard those sent and measured in different bases.
7. Together, they can reconstitute the random bit string created previously by Alice.

Here 1$^{st}$ stage of BB84 is completed. Now Bob and Alice have the common shared key that is perfectly secure. In 1$^{st}$ stage Bob got 8 bits matched out of 16 bits. The process of 1$^{st}$ stage can be shown with the help of Table 2.

| 1$^{st}$ Stage (Transmission from Alice to Bob) | | | | | | | | | | | | | | | | |
|---|---|---|---|---|---|---|---|---|---|---|---|---|---|---|---|---|
| Alice | 1 | 1 | 0 | 0 | 0 | 1 | 1 | 0 | 1 | 0 | 0 | 1 | 0 | 0 | 1 | 1 |
| Logic circuit output | 1 | 1 | 0 | 0 | 1 | 0 | 0 | 1 | 0 | 1 | 1 | 0 | 0 | 0 | 1 | 1 |
| Basis | X | + | + | X | + | + | X | X | X | X | + | + | + | X | X | + |
| Polarization | ↗ | ↑ | → | ↖ | ↑ | → | ↘ | ↘ | ↗ | ↑ | → | → | ↖ | ↗ | ↑ |
| B's Basis | + | + | + | X | X | + | X | X | X | X | X | + | + | X | + | + |
| Polarization | ↑ | → | → | ↖ | ↗ | ↑ | ↘ | ↖ | ↗ | ↖ | ↑ | → | ↘ | ↑ | → |
| Measurement of bits | 1 | 0 | 0 | 0 | 1 | 1 | 0 | 1 | 0 | 1 | 1 | 1 | 0 | 0 | 1 | 0 |
| Matching pairs | - | - | 0 | 0 | - | - | 0 | 1 | 0 | 1 | - | - | 0 | 0 | - | - |
| Inverse logic circuit | - | - | 0 | 0 | - | - | 1 | 0 | 1 | 0 | - | - | 0 | 0 | - | - |
| 1$^{st}$ stage shared key | - | - | 0 | 0 | - | - | 1 | 0 | 1 | 0 | - | - | 0 | 0 | - | - |

Table 2. Transmission from Alice to Bob

### 4.2.2 Second Stage from Bob to Alice

In 1$^{st}$ stage Bob got 8 bits matched out of 16 bits. In 2$^{nd}$ stage I have tried to increase this number so that accuracy relative to the original bit sequence decided by Alice can be achieved in a better way.
1. Bob decides the Source Basis only for those bits that did not match and applies the polarization. Here in this example: X+--++----++--X+.
2. Alice puts her predefined basis for measurement and applies the polarization.
3. Then Alice does measurement of bit combination that is common. Final secret key after second stage can be taken as any subset of the final output. For example: I have taken last bit of every combination i.e. 100001.

These combinations of bits are further added to the 1st stage shared keys. Final number of bits matched is 12 out of 16 bits in the above example.

The process of 2nd stage can be shown with the help of table 3:

| 2nd Stage (transmission from Bob to Alice) | | | | | | | | | | | | | | | |
|---|---|---|---|---|---|---|---|---|---|---|---|---|---|---|---|
| Bob Random Basis | X | + | - | - | + | + | - | - | - | - | + | + | - | - | X | + |
| Polarization | ↗ | ↑ | - | - | ↑ | → | - | - | - | - | ↑ | → | - | - | ↗ | ↑ |
| Measurement of bits by A | 1 | 1 | - | - | 1 | 0 | - | - | - | - | 1 | 0 | - | - | 1 | 1 |
| Matching pairs | 1 | 1 | - | - | - | - | - | - | - | - | - | - | - | - | 1 | 1 |
| 1st shared key | - | - | 0 | 0 | - | - | 1 | 0 | 1 | 0 | - | - | 0 | 0 | - | - |
| 2nd shared key | 1 | 1 | - | - | - | - | - | - | - | - | - | - | - | - | 1 | 1 |
| Final Shared key | 1 | 1 | 0 | 0 | - | - | 1 | 0 | 1 | 0 | - | - | 0 | 0 | 1 | 1 |

Table 3. Transmission from Bob to Alice

### 4.3 Comparative analysis

**BB84 protocol**

Suppose if we have 2880 bits. If we apply probability then in 1440 bits rectilinear base will be applied by Alice and in other 1440 bits diagonal base will be applied. On an average if Eve intercepts the photons then in ½ the times random base used by her, will be same as used by Alice. i.e. 1440 bits. Now Bob will also be using the same random bases as by Alice. So, on an average 720 of the bits used by Alice, bob and Eve will be same. So this indicates that out of 1440 bits key 720 bits will be known to Eve.
Conclusion:
Total bits used=2880
No of bits used for formation of key=1440
No of bits that Eve can guess=720
i.e. Eve knows half of the key.

**Proposed Technique**

Proposed technique combines pairs of bits therefore there are 4 possible combinations: X++XXX++. Suppose we have 2880 bits i.e. 1440 bits pair. Here if we intercept the photons then in 1/4 of times the random bases used i.e. 380 will be same by Alice because there are 4 possible combinations X++XXX++. So there are 380 combinations of bits that Bob and Alice matches. Further 90 combinations out of 380 are matches for Eve, Alice and Bob.

Conclusion:
Total no of bits used= 2880 or 1440 combinations of 2 bits each.
No. of bits used for the formation of key=380
No of bits Eve can guess=90

i.e Eve knows the ¼ of the key.

The statistics can be explained better with the help of comparison table:

| Sr. No. | Criteria | BB84 Protocol | Proposed Technique |
|---|---|---|---|
| 1 | No of Bits | 2880 | 2880 |
| 2 | Combination | 1440 pairs | 1440 combinations of 2 bits each |
| 3 | No of bits used for key formation | 1440 | 360 |
| 4 | No of bits that Eve can guess | 720 | 90(4 possible combinations) |
| 5 | Security | 50% | 75% |

Table 4: Comparison Table

### 3.4 Information Reconciliation and Privacy Amplification

The proposed technique uses double BB84 protocol. The process of reconciliation results in a bit sequence which is common to Alice and Bob, but some of its bits may be known to an eavesdropper who has tapped the classical channel. To eliminate this "leaked" information, Alice and Bob must apply, in common, a binary transformation (usually, a random permutation) to their sequences, and discard a subset of bits from the result. [13] Privacy amplification uses Alice and Bob's key to produce a new key, in such a way that Eve has only negligible information about the new key. One of the methods to do this is using a universal hash function, chosen at random from a publicly known set of such functions, which takes as its input a binary string of length equal to the key and outputs a binary string of a chosen shorter length. Information reconciliation is applied in this technique using the BB84 protocol at two levels. This information is known to Alice and Bob because the input to the $2^{nd}$ BB84 protocol is completely random and hence Eve can not intercept any information, this helps in privacy amplification. [3]

## 4. CONCLUSION

The proposed technique can be used as a powerful tool for solving the problems related to quantum key distribution and provides better security than previous BB84 protocol because this technique uses the BB84 protocol twice along with the logic circuit at the first stage to enhance the security level. Comparison table shows that the BB84 protocol provides 50% security whereas the proposed technique provides 75% it means the proposed technique is more efficient and provides 25 % more security than the BB84 protocol. Hence we can improve the efficiency and performance to an extent of BB84 protocol by using the BB84 protocol twice along with the logic circuit but at the cost of complexity. The security level can be increase further by using the concept of information reconciliation and privacy amplification.

**Author1**

Dr. Ajit Singh is presently working as Reader and Head of CSE & IT Department of School of Engineering & Sciences in BPSMV, Khanpur Kalan (Sonepat). He is also having the additional charge as a Director of University Computer Center (UGC). He posses qualifications of B.Tech, M.Tech, Ph.D (p). He is a member of BOG (Board of Governors) of Haryana State Counselling Society, Panchkula and also member of academic council in the University. He published approximate 20 papers in National/ International journals and conferences and holds a teaching experience of approximate 10 years. He holds the membership of Internal Quality Assurance cell, UG-BOS & PG-BOS and the NSS advisory committee. He is also an associate member of CSI & IETE. His research interests are in Network Security, Computer Architecture and Data Structure.

**Author2**

Ms. Nidhi Sharma has completed her B.Tech degree in Information Technology from Vaish College of Engineering, Rohtak, Maharshi Dayanand University (MDU), India in the year 2009, and She is pursuing M.Tech in Information Technology, Bansathali university Banasthali from June 2009. Currently she is Doing Internship from B.P.S.M.V Khanpur Kalan, Sonipat. She has presented papers in International and National Conferences. Her research Interests are in Network Secuirty and Software engineering.